\documentclass[twocolumn]{aa}
\usepackage{graphicx}
\usepackage{txfonts}
\usepackage{natbib}

\begin{document}
   \title{XMM-Newton Detection of Hot Gas in the Eskimo Nebula: 
          Shocked Stellar Wind or Collimated Outflows?
\thanks{Based on observations obtained with \emph{XMM-Newton}, 
 an ESA science mission with instruments and contributions directly funded
 by ESA Member States and NASA.}}

   \titlerunning{Hot Gas in the Eskimo Nebula}
   

   \author{M.\ A.\ Guerrero\inst{1},
          Y.-H. Chu\inst{2},
          R.\ A. Gruendl\inst{2}, and 
          M.\ Meixner\inst{3}
          }

   \authorrunning{M.~A.\ Guerrero et al.}

   \offprints{M.~A.\ Guerrero}

   \institute{Instituto de Astrof\'{\i}sica de Andaluc\'{\i}a, CSIC,
            Apartado Correos 3004, E-18080, Granada, Spain \\
            \email{mar@iaa.es}
        \and
            Astronomy Department, University of Illinois,
            1002 W. Green Street, Urbana, IL 61801, USA  \\
            \email{chu@astro.uiuc.edu, gruendl@astro.uiuc.edu}
        \and
            Space Telescope Science Institute, 3700 San Martin Drive, 
            Baltimore, MD 21218, USA   \\
            \email{meixner@stsci.edu}
             }

   \date{Received November 2004 / Accepted later 2004}

   \abstract{
The Eskimo Nebula (NGC\,2392) is a double-shell planetary nebula (PN)
known for the exceptionally large expansion velocity of its inner shell,
$\sim$90 km~s$^{-1}$, and the existence of a fast bipolar outflow with a
line-of-sight expansion velocity approaching 200 km~s$^{-1}$. 
We have obtained \emph{XMM-Newton} observations of the Eskimo and 
detected diffuse X-ray emission within its inner shell. 
The X-ray spectra suggest thin plasma emission with a temperature 
of $\sim2\times10^6$ K and an X-ray luminosity of $L_{\rm X} = 
(2.6\pm1.0)\times10^{31} (d/1150~{\rm pc})^2$ ergs~s$^{-1}$, where 
$d$ is the distance in parsecs.
The diffuse X-ray emission shows noticeably different spatial
distributions between the 0.2--0.65 keV and 0.65--2.0 keV bands.
High-resolution X-ray images of the Eskimo are needed to determine 
whether its diffuse X-ray emission originates from shocked fast wind 
or bipolar outflows.  
   }

   \maketitle
%

\section{Introduction}

Planetary nebulae (PNe) consist of the stellar material ejected 
by low- and intermediate-mass stars (1--8 M$_\odot$) at the end 
of the asymptotic giant branch (AGB) phase.  
As such a star evolves off the AGB, the copious mass-loss strips 
off the stellar envelope and exposes the hot stellar core.  
A PN emerges when the stellar UV radiation ionizes the ejected 
stellar material, causing it to emit in the optical.

PNe are expected to be diffuse X-ray sources.
The central stars of PNe possess fast stellar winds with terminal 
velocities of 1000--4000 km~s$^{-1}$ \citep{CP85}, while fast 
collimated outflows with velocities up to 1000 km~s$^{-1}$ are 
also observed in PNe and proto-PNe, e.g., MyCn\,18 and Hen\,3-1475 
\citep{Betal95,Retal95,OCetal00}.  
The interactions of the fast stellar wind and/or collimated outflows 
with nebular material produce shocked gas that is hot enough to
emit in X-rays.

In the interacting-stellar-winds model of PNe \citep{KPF78}, 
the fast wind emanating from the central star sweeps up the slow 
AGB wind to form a sharp nebular shell.
The interior structure of a PN would be similar to that of a 
wind-blown bubble \citep[e.g.,][]{Wetal77}.  
The central cavity of a PN is expected to be filled with 
shocked fast wind at 10$^7$--10$^8$~K, but this hot gas is too 
tenuous to produce appreciable X-ray emission.  
Dynamic or evaporative mixing of cool nebular material into the hot 
gas at their interface produces optimal conditions for soft X-ray 
emission, which will show a limb-brightened morphology within the 
nebular shell, as observed in NGC\,6543 \citep{Cetal01}.

Fast collimated outflows or jets may also produce hot X-ray-emitting 
gas.  
When outflows with velocities $\ge$300~km~s$^{-1}$ initially impinge 
on the AGB wind, bow shocks and X-ray emission can be produced 
 \citep[e.g., Hen\,3-1475,][]{Setal03}.  
The prolonged action of collimated outflows may bore through the
AGB wind and form extended cavities filled by shocked hot gas 
that emits X-rays \citep[e.g., Mz\,3,][]{Ketal03}.

Diffuse X-ray emission from hot gas in PNe was hinted by \emph{ROSAT}
observations \citep{GCG00}, but was unambiguously resolved only by 
\emph{Chandra} and \emph{XMM-Newton} observations.
Besides the aforementioned Hen\,3-1475, Mz\,3, and NGC\,6543, diffuse 
X-ray emission has been reported in only 4 other PNe: A\,30, 
BD+30$^\circ$3639, NGC\,7009, and NGC\,7027 
\citep{CCC97,GGC02,Ketal00,KVS01}.  
It is imperative to detect diffuse X-ray emission from a large number
of PNe to investigate whether shocked fast stellar wind, collimated 
outflows or both are responsible for the X-ray emission from PNe.

\begin{figure*}[!t]
\centering
\includegraphics[width=18.5cm]{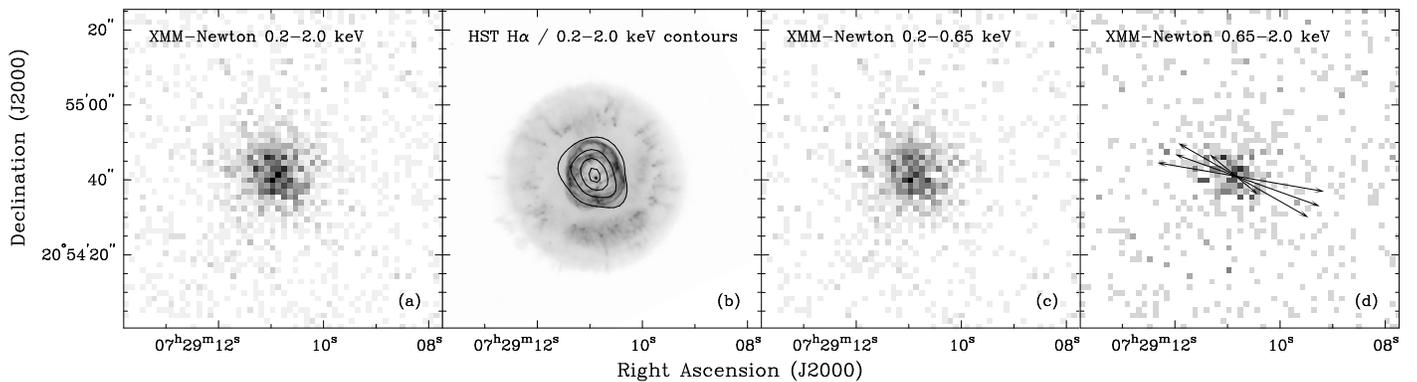}
\caption{
\emph{XMM-Newton} and \emph{HST} H$\alpha$ images of the Eskimo Nebula. 
Panel {\it (a)} displays the \emph{XMM-Newton} EPIC raw image in the 
0.2-2.0 keV band with pixel size 1\farcs5. 
Panel {\it (b)} shows the \emph{HST} H$\alpha$ image overplotted by 
the 25\%, 50\%, 75\%, and 95\% X-ray contours extracted from the 
adaptively smoothed 0.2-2.0 keV band image described in the text.
Contours at fainter levels are not plotted, as they show mainly the 
extended wings of \emph{XMM-Newton}'s PSF and the smear caused by 
the adaptive smoothing.  
Panels {\it (c)} and {\it (d)} display \emph{XMM-Newton} EPIC raw 
images in the 0.2--0.65 and 0.65--2.0 keV energy bands, respectively. 
The pixel size in these images is 1\farcs5.  
The arrows in panel {\it (d)} mark the location of the fast collimated 
outflow as derived from high-dispersion echelle spectroscopy 
\citep[][Maestro et al., in preparation]{GBS85}. 
}
\label{Images}
\end{figure*}

The Eskimo Nebula (NGC\,2392) offers an excellent opportunity to assess 
the effects of collimated outflows in a PN.  
The Eskimo is a double-shell PN: its bright elliptical inner shell has 
two blister-like protrusions on the south rim, and its round outer shell
contains a set of complex low-ionization features \citep{ODetal02}.  
In addition, the Eskimo has a fast collimated outflow, detected in
high-dispersion spectra of optical nebular lines; the outflow, with 
line-of-sight expansion velocities approaching 200~km~s$^{-1}$, is 
dynamically interacting with nebular material \citep[][Maestro et al., 
in preparation]{GBS85,ODB85,OWC90}.
We have obtained \emph{XMM-Newton} observations of the Eskimo and here
we report the detection of diffuse X-ray emission from this nebula.

\section{Observations}

The Eskimo Nebula was observed with the \emph{XMM-Newton Observatory} 
in Revolution 790 on 2004 April 2 using the EPIC/MOS1, EPIC/MOS2, 
and EPIC/pn CCD cameras.  
The two EPIC/MOS cameras were operated in the Full-Frame Mode for a 
total exposure time of 17.5 ks, while the EPIC/pn camera was operated 
in the Extended Full-Frame Mode for a total exposure time of 12.5 ks.  
For all observations, the Medium filter was used.  
The \emph{XMM-Newton} pipeline products were processed using the 
\emph{XMM-Newton} Science Analysis Software (SAS version 6.0.0) and 
the calibration files from the Calibration Access Layer available 
on 2004 June 10.

The event files were screened to eliminate events due to charged 
particles or associated with periods of high background.  
For the EPIC/MOS observations, only events with CCD patterns 0--12 
(similar to $ASCA$ grades 0--4) were selected;  
for the EPIC/pn observation, only events with CCD pattern 0 (single 
pixel events) were selected.  
Time intervals with high background (i.e.\ count rates 
$\ge 0.3$ cnts~s$^{-1}$ for the EPIC/MOS or $\ge 1.4$ cnts~s$^{-1}$ 
for the EPIC/pn in the background-dominated 10--12 keV energy range)
were discarded.
The resulting exposure times are 17.3 ks, 17.1 ks, and 11.6 ks for 
the EPIC/MOS1, EPIC/MOS2, and EPIC/pn observations, respectively.

\section{Results}

The \emph{XMM-Newton} EPIC/MOS1, EPIC/MOS2, and EPIC/pn observations 
of the Eskimo detect within the nebula a total of $180\pm15$, 
$175\pm15$, and $620\pm30$ cnts, respectively.  
In order to construct an X-ray image of the highest signal-to-noise 
ratio, we merged together the event files of the three EPIC
observations and extracted raw EPIC images in the 0.2--2.0, 0.2--0.65, 
and 0.65--2.0 keV bands with a pixel size of 1\farcs5 (Fig.~1a, 1c and 
1d).  
The raw EPIC image in the 0.2--2.0 keV band is then adaptively 
smoothed using Gaussian profiles with FWHM ranging from 4$''$ to 6$''$.
The contour map of this smoothed image is overplotted on a  
\emph{Hubble Space Telescope (HST)} Wide Field Planetary Camera 2 
(WFPC2) H$\alpha$ image (Fig.~1b).  
The alignment of X-ray and optical images is fine-tuned using 
HD\,59087, a star $\sim$100$''$ north of the Eskimo, which is 
detected both in the \emph{XMM-Newton} EPIC and \emph{HST} WFPC2 
images.

The X-ray emission from the Eskimo is clearly extended.  
Its distribution is elongated along PA$\sim$25$^\circ$ (Fig.1a), 
and the 25\% contour of the smoothed EPIC image follows closely 
the outline of the the inner shell in the \emph{HST} H$\alpha$ 
image (Fig.~1b).  
The image in the 0.2--0.65 keV band shows a similar spatial 
distribution, but in the 0.65--2.0 keV band the emission is 
elongated along PA$\sim$70$^\circ$, i.e., roughly aligned 
with the fast bipolar outflow detected at PAs of 50$^\circ$--80$^\circ$
and 230$^\circ$--260$^\circ$, as illustrated in Fig.~1d 
\citep[][Maestro et al., in preparation]{GBS85}. 
These comparisons suggest that the diffuse X-ray emission from the
Eskimo is mostly confined within its inner shell, but some of the 
harder X-ray emission in the 0.65--2.0 keV band may be produced by 
the interaction of the fast bipolar outflow with nebular material.

\begin{figure}[!t]
\resizebox{\hsize}{!}{\includegraphics{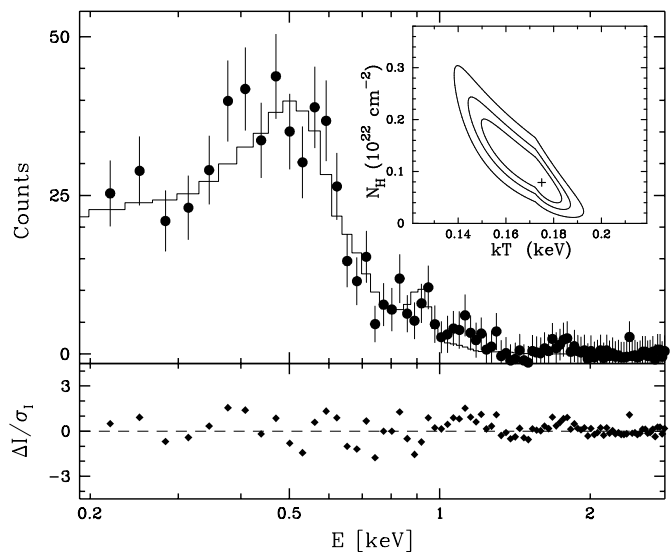}}
\caption{
{\it (top)} 
\emph{XMM-Newton} EPIC/pn background-subtracted spectrum of the Eskimo 
overplotted with the best-fit MEKAL model.  
For plotting purposes, both the spectrum and model are binned in 30 eV 
wide energy bins.  
{\it (inset)} $\chi^2$ grid plot as a function of $kT$ and $N_{\rm H}$ 
of the spectral fit to the EPIC/pn spectrum.  
The contours represent 68\%, 90\%, and 99\% confidence levels. 
{\it (bottom)} Relative residuals of the best-fit model to the EPIC/pn 
spectrum.  
}
\label{Spectrum}
\end{figure}

To analyze the properties of the diffuse X-ray emission from the Eskimo,
we extracted a spectrum from the EPIC/pn event file using a circular 
source aperture of 20$''$ radius, large enough to include all X-ray
emission from the nebula.
The background level was assessed from an annular region exterior to
the source aperture and 9 times larger in area.
The background-subtracted EPIC/pn spectrum of the Eskimo is shown in 
Fig.~2.  
This spectrum shows a broad peak between 0.4 and 0.6 keV,
a plateau below 0.4 keV, and a sharp decline above 0.6 keV.
Low-intensity emission peaks are discernible at $\sim$0.9 keV, 
$\sim$1.1 keV, and $\sim$1.8 keV.  
The overall spectral shape indicates thermal plasma emission, and 
the peaks correspond to the He-like triplets of 
N~{\sc vi}    at $\sim$0.43 keV, 
O~{\sc vii}   at $\sim$0.57 keV, 
Ne~{\sc ix}   at $\sim$0.92 keV, 
Na~{\sc x}    at $\sim$1.13 keV, and 
Si~{\sc xiii} at $\sim$1.86 keV. 
We have also extracted EPIC/MOS spectra; they show similar spectral
features at lower signal-to-noise ratios, so we will concentrate
only on the EPIC/pn spectrum in the analysis below.

The observed spectrum can be modeled to determine the physical 
conditions of the X-ray-emitting gas and the amount of foreground 
absorption.  
We have adopted the MEKAL optically-thin plasma emission model 
\citep{KM93,LOG95} and nebular abundances, 
although we note that these may differ from those of the 
X-ray-emitting gas \citep[e.g.,][]{Metal03}.  
The abundances of He, C, N, O, Ne, Ar, and S relative to
hydrogen by number, as determined from optical and UV spectra
of the Eskimo, are 0.076, 2.2 $\times$ 10$^{-4}$, $1.1\times10^{-4}$,
$2.8\times10^{-4}$, $6.4\times10^{-5}$, $1.4\times10^{-6}$, and 
$4.3\times10^{-6}$, respectively \citep{B91,HKB00}. 
For elements that do not have available nebular abundances, we adopt 
solar abundances \citep{GS98}.  
We have also assumed solar abundances for the foreground interstellar 
absorption, and adopted absorption cross-sections from \citet{BM92}.

The spectral fit is carried out by folding the absorbed MEKAL model 
spectrum through the EPIC/pn response matrix, and comparing the 
modeled spectrum to the observed EPIC/pn spectra in the 0.2--2.5 keV 
energy range using the $\chi^2$ statistics.  
The best-fit model with nebular abundances has a large reduced $\chi^2$, 
$\sim$2.5, with excessive positive residuals at $\sim$0.4 keV and 
$\sim$0.9 keV and negative residuals at $\sim$0.6 keV.  
These energies are coincident with the N~{\sc vi}, Ne~{\sc ix}, and
O~{\sc vii} lines, respectively; thus the residuals suggest that 
the N/O and Ne/O abundance ratios are higher than those of the
adopted nebular values.
We have therefore made spectral fits that allowed the abundances of N
and Ne to vary, and subsequently the reduced $\chi^2$ is improved to 
$\sim$1.5.

The best-fit model, overplotted on the EPIC/pn spectrum in Fig.~2, has 
a plasma temperature of $T = (2.0^{+0.1}_{-0.3}) \times10^6$ K (or 
$kT$ = 0.175$^{+0.01}_{-0.03}$ keV), an absorption column density 
$N_{\rm H} = (8^{+12}_{-5})\times10^{20}$ cm$^{-2}$, and a volume 
emission measure of (1.2$\pm$0.3)$\times$10$^{54} (d/1150 {\rm pc})^2$ 
cm$^{-3}$, where $d$ is the distance in parsecs and has been reported 
to be 1150 pc \citep{Petal97}.  
The quality of the spectral fits is illustrated by the plot of 
the reduced $\chi^2$ of the fits as a function of $kT$ and 
$N_{\rm H}$ shown in the inset of Fig.~2.  
Adopting a typical gas-to-dust ratio \citep{BSD78}, the best-fit 
absorption column density corresponds to a logarithmic extinction at 
the H$\beta$ line of $c_{\rm H\beta}$ = $0.2^{+0.3}_{-0.1}$, in 
agreement with that determined from the Balmer decrement \citep{B91}.  
The N and Ne abundances are enhanced with respect to the nebular 
abundances by 3.5$\pm$1.2 and 3.0$\pm$1.0, respectively, thus 
resulting in N/O$\sim$1.4 and Ne/O$\sim$0.7.  
The residuals of the best-fit model may also indicate enhanced Na and 
Si abundances, but the number of counts is too small to warrant 
spectral fits allowing these abundances to vary.

The observed (absorbed) X-ray flux in the 0.2--2.5 keV energy range 
is $(6\pm1)\times10^{-14}$ ergs~cm$^{-2}$~s$^{-1}$, 
and the intrinsic (unabsorbed) X-ray flux is 
$(1.6\pm0.6)\times10^{-13}$ ergs~cm$^{-2}$~s$^{-1}$. 
The X-ray luminosity in the same energy range is $L_{\rm X} = 
(2.6\pm1.0)\times10^{31} (d/1150~{\rm pc})^2$ ergs~s$^{-1}$.

\section {Discussion}

The sharp rim of the Eskimo's inner shell suggests compression by
supersonic shocks.
Indeed, the inner shell of the Eskimo,  expanding at $\sim$90 km~s$^{-1}$ 
into the much slower outer shell, has an expansion velocity notably high 
among PNe \citep{ODB85}.  
To assess whether the hot gas in the central cavity is responsible
for driving such a fast expansion, we derive the thermal pressure 
of the X-ray-emitting gas and compare it to those of the surrounding 
nebular shells.  
Assuming a prolate ellipsoidal central cavity, the volume occupied 
by the X-ray-emitting gas is $\sim1.1\times10^{51} (d/1150~{\rm pc})^3 
(\epsilon/0.5)$ cm$^3$, where $\epsilon$ is the filling factor and 
may be $\sim$0.5.  
From this volume and the aforementioned volume emission measure, 
we derive an rms electron density of  $n_{\rm e}
\sim$35~$(d/1150 {\rm pc})^{-1/2}~(\epsilon/0.5)^{-1/2}$ cm$^{-3}$.  
This rms electron density and the plasma temperature of $2.0\times10^6$~K 
imply that the thermal pressure of the X-ray-emitting gas, 
$P_{\rm th} \sim 1.9 \times n_{\rm e} k T$, is 
$\sim2\times10^{-8}~(d/1150 {\rm pc})^{-1/2}~(\epsilon/0.5)^{-1/2}$ 
dynes~cm$^{-2}$.
The inner and outer shells of the Eskimo have densities of 2500 and 
900 cm$^{-3}$, respectively, and a temperature of 14,500 K \citep{B91}.  
The thermal pressure of the inner and outer shells are 
$\sim$1$\times 10^{-8}$ and $\sim$3$\times 10^{-9}$ dynes~cm$^{-2}$, 
respectively.  
The X-ray-emitting gas has slightly higher thermal pressure than the 
inner shell, and much higher than the outer shell, thus  
the thermal pressure of the hot gas in the Eskimo's central
cavity drives the expansion of its inner shell into the outer shell. 

The temperature of the X-ray-emitting gas in the Eskimo, 
2.0$\times$10$^6$~K, is similar to that found in other elliptical PNe;
however, the terminal velocity of the Eskimo's fast wind is low,
only $\sim$400 km~s$^{-1}$ \citep{PHM02}.
For an adiabatic shock, the expected temperature of the shocked wind 
would be $\sim$1.9$\times$10$^6$~K for such a velocity, barely 
reaching the temperature indicated by the X-ray spectrum, in sharp 
contrast to other PNe whose shocked stellar winds are 10--100 times 
hotter than the hot gas detected \citep[e.g.,][]{Cetal01,Ketal00}.  
If the hot gas in the Eskimo indeed originates from the shocked
fast stellar wind, the low expected post-shock temperature implies 
that no significant mixing of nebular material has taken place.
The mass of the hot gas, 
$\sim$3$\times10^{-5}~(d/1150 {\rm pc})^{5/2}~(\epsilon/0.5)^{1/2}$ M$_\odot$,
can be supplied by the fast stellar wind in $\sim$1800 yr at a 
constant mass loss rate of 1.8$\times$10$^{-8}$ M$_\odot$~yr$^{-1}$
\citep{PHM02}.

If the X-ray emission from the Eskimo does not originate 
from a shocked fast stellar wind, then there are two other possibilities.
First, the X-ray emission can be partially attributed to the central 
star as in the case of NGC\,6543 or to a late-type binary companion 
as in the case of NGC\,7293 \citep{Getal01}.  
The angular resolution of \emph{XMM-Newton}, however, is insufficient 
to resolve a point source from the diffuse emission from the Eskimo.
Second, the diffuse X-ray emission can be produced by the dynamical
interaction of the fast bipolar outflow with the inner shell.  
This collision will produce shock-excited gas along the direction 
of the outflow which is detected as the harder X-ray emission in 
the 0.65--2.0 keV band, and also as shock-excited [Fe~{\sc ii}] 
1.26 and 1.64 $\mu$m line emissions \citep{HM89,HLD99}.  
Similar situation is observed in BD+30$^\circ$3639, whose diffuse 
X-ray emission shows an asymmetric spatial distribution in hard 
energies and reveals spectral evidence of enhanced Ne/O abundance 
ratio \citep{Ketal02}. 
As suggested for BD+30$^\circ$3639 \citep{SK03}, the hottest gas 
in the Eskimo may have been produced by a fast, $\sim$500 km~s$^{-1}$, 
collimated post-AGB wind that is no longer present.  
Alternatively, the observed 200 km~s$^{-1}$ outflow may have a 
large inclination with respect to the line-of-sight, so that its 
real expansion velocity is large enough to power the hottest gas 
in the Eskimo.

Our \emph{XMM-Newton} observations of the Eskimo Nebula have detected 
diffuse X-ray emission from hot gas within its central cavity.  
The high thermal pressure of this hot gas is responsible for the 
high expansion velocity of its innermost shell.  
The origin of this hot gas, however, is uncertain, and may
consist of several components: shocked fast stellar wind, 
shocks associated with the fast bipolar outflow, and emission
from its central star or a binary companion.  
X-ray observations at the highest spatial resolution afforded
by \emph{Chandra} are needed to determine accurately the origin 
of X-ray emission from the Eskimo Nebula, a very interesting and
complex PN.

\begin{acknowledgements}
M.A.G. is grateful to the VILSPA staff for his help and hospitality 
during the 4$^{\rm th}$ SAS Workshop held in VILSPA on June 8-12 2004.  
M.A.G. also acknowledges support from the grant AYA~2002-00376 of 
the Spanish MCyT (cofunded by FEDER funds). Y.-H.C. acknowledges 
support from the NASA grant NNG04GE63G. 
\end{acknowledgements}


\begin{thebibliography}{}

\bibitem[Balucinska-Church \& McCammon (1992)]{BM92} 
         Balucinska-Church, M., \& McCammon, D.\ 1992, ApJ, 400, 699

\bibitem[Barker(1991)]{B91}
Barker, T.\ 1991, \apj, 371, 217

\bibitem[Bohlin, Savage, \& Drake(1978)]{BSD78}
Bohlin, R.~C., Savage, B.~D., \& Drake, J.~F.\ 1978, \apj, 224, 132

\bibitem[Bobrowsky et al.(1995)]{Betal95} 
Bobrowsky, M.\ et al.\ 1995
\apjl, 446, L89 

\bibitem[Cerruti-Sola \& Perinotto(1985)]{CP85} 
Cerruti-Sola, M.~\& Perinotto, M.\ 1985, \apj, 291, 237 

\bibitem[Chu, Chang, \& Conway(1997)]{CCC97} 
Chu, Y.-H., Chang, T.~H., \& Conway, G.~M.\ 1997, \apj, 482, 891 

\bibitem[Chu et al.(2001)]{Cetal01} 
Chu, Y.-H., Guerrero, M.~A., Gruendl, R.~A., Williams, R.~M., \& Kaler, 
J.~B.\ 2001, \apjl, 553, L69 

\bibitem[Gieseking, Becker, \& Solf(1985)]{GBS85}
Gieseking, F., Becker, I., \& Solf, J.\ 1985, \apjl, 295, L17

\bibitem[Grevesse \& Sauval(1998)]{GS98} 
         Grevesse, N., \& Sauval, A.\ J.\ 1998, \ssr, 85, 161

\bibitem[Guerrero, Chu, \& Gruendl(2000)]{GCG00} 
Guerrero, M.~A., Chu, Y.-H., \& Gruendl, R.~A.\ 2000, \apjs, 129, 295 

\bibitem[Guerrero et al.(2001)]{Getal01} 
Guerrero, M.~A., Chu, Y.-H., Gruendl, R.~A., Williams, R.~M., \& Kaler, 
J.~B.\ 2001, \apjl, 553, L55

\bibitem[Guerrero, Gruendl, \& Chu(2002)]{GGC02} 
Guerrero, M.~A., Gruendl, R.~A., \& Chu, Y.-H.\ 2002, \aap, 387, L1 

\bibitem[Henry, Kwitter \& Bates(2000)]{HKB00}
Henry, R.\ B.\ C., Kwitter, K.\ B., \& Bates, J.\ A.\ 2000, \apj, 531, 928

\bibitem[Hollenbach \& McKee(1989)]{HM89}
Hollenbach, D.~\& McKee, C.~F.\ 1989, \apj, 342, 306 

\bibitem[Hora, Latter, \& Deutsch(1999)]{HLD99}
Hora, J.~L., Latter, W.~B., \& Deutsch, L.~K.\ 1999, \apjs, 124, 195 

\bibitem[Kaastra \& Mewe(1993)]{KM93}
Kaastra, J.\ S., \& Mewe, R.\ 1993, Legacy, 3, 16, HEASARC, NASA

\bibitem[Kastner et al.(2003)]{Ketal03} 
Kastner, J.~H., Balick, B., Blackman, E.~G., Frank, A., Soker, N., 
Vrt{\'{\i}}lek, S.~D., \& Li, J.\ 2003, \apjl, 591, L37 

\bibitem[Kastner et al.(2002)]{Ketal02} 
Kastner, J.~H., Li, J., Vrtilek, S.~D., Gatley, I., Merrill, K.~M., \& 
Soker, N.\ 2002, \apj, 581, 1225

\bibitem[Kastner et al.(2000)]{Ketal00} 
Kastner, J.~H., Soker, N., Vrtilek, S.~D., \& Dgani, R.\ 2000, \apjl, 545, 
L57 

\bibitem[Kastner, Vrtilek, \& Soker(2001)]{KVS01} 
Kastner, J.~H., Vrtilek, S.~D., \& Soker, N.\ 2001, \apjl, 550, L189 

\bibitem[Kwok, Purton, \& Fitzgerald(1978)]{KPF78} 
Kwok, S., Purton, C.~R., \& Fitzgerald, P.~M.\ 1978, \apjl, 219, L125 

\bibitem[Liedhal, Osterheld, \& Goldstein(1995)]{LOG95}
Liedahl, D.\ A., Osterheld, A.\ L., \& Goldstein, W.\ H.\ 1995, 
\apj, 438, L115

\bibitem[Maness et al.(2003)]{Metal03}
Maness, H.\ L., Vrtilek, S.\ D., Kastner, J.\ H., \& Soker, N.\ 2003, 
\apj, 589, 439 

\bibitem[O'Connor et al.(2000)]{OCetal00} 
O'Connor, J.~A., Redman, M.~P., Holloway, A.~J., Bryce, M., L{\' o}pez, 
J.~A., \& Meaburn, J.\ 2000, \apj, 531, 336 

\bibitem[O'Dell et al.(2002)]{ODetal02} 
O'Dell, C.~R., Balick, B., Hajian, A.~R., Henney, W.~J., \& Burkert, A.\ 
2002, \aj, 123, 3329 

\bibitem[O'Dell \& Ball(1985)]{ODB85} 
O'Dell, C.~R.~\& Ball, M.~E.\ 1985, \apj, 289, 526 

\bibitem[O'Dell, Weiner, \& Chu(1990)]{OWC90} 
O'Dell, C.~R., Weiner, L.~D., \& Chu, Y.-H.\ 1990, \apj, 362, 226 

\bibitem[Pauldrach, Hoffmann, \& M{\' e}ndez(2003)]{PHM02} 
Pauldrach, A.~W.~A., Hoffmann, T.~L., \& M{\' e}ndez, R.~H.\ 2003, IAU 
Symposium, 209, 177 

\bibitem[Perryman et al.(1997)]{Petal97}
Perryman, M.~A.~C.\ et al., 1997, \aap, 323, L49

\bibitem[Riera et al.(1995)]{Retal95} 
Riera, A., Garc\'{\i}a-Lario, P., Manchado, A., Pottasch, S.~R., 
\& Raga, A.~C.\ 1995, \aap, 302, 137 

\bibitem[Sahai et al.(2003)]{Setal03} 
Sahai, R., Kastner, J.~H., Frank, A., Morris, M., \& Blackman, E.~G.\ 
2003, \apjl, 599, L87 

\bibitem[Soker \& Kastner(2003)]{SK03} 
Soker, N.~\& Kastner, J.~H.\ 2003, \apj, 583, 368 

\bibitem[Weaver et al.(1977)]{Wetal77} 
Weaver, R., McCray, R., Castor, J., Shapiro, P., \& Moore, R.\ 1977, 
\apj, 218, 377 

\end{thebibliography}
\end{document}